\documentstyle[12pt]{article}
\def\ffrac#1#2{\textstyle{#1\over#2}\displaystyle}
\def\bm#1{\mbox{\boldmath $#1$}}

\begin{document}

\begin{center}
{\bf \LARGE Ground state energy and mass gap of a\\ 
generalised quantum spin ladder\\}  
\vspace{8mm}
{\bf \large M. T. Batchelor and M. Maslen\\}
\vspace{2mm}
{Department of Mathematics,
School of Mathematical Sciences,\\
Australian National University, Canberra ACT 0200, Australia\\}
\end{center}
\vspace{6mm}
\begin{abstract}
We show that a 2-leg ladder hamiltonian introduced recently by
Albeverio and Fei can be made to satisfy the Hecke algebra.
As a result we have found an equivalent representation of the 
eigenspectrum in terms of the spin-$\ffrac12$ antiferromagnetic
XXZ chain at $\Delta = -\ffrac53$. The values of thermodynamic
quantities such as the ground state energy and mass gap follow 
from the known XXZ results.
\end{abstract}
\newpage

\section{Introduction}

A number of solvable models of quantum spin ladders have recently been
found. In some models certain combinations of couplings exist such that
ground states can be constructed in simple form \cite{KM}. 
Other models have an underlying $R$-matrix and are thus amenable
to the machinery of exactly solvable lattice models in statistical
mechanics. Of particular relevence here are the 2-leg Heisenberg ladders of 
Albeverio and Fei \cite{AF} and Wang \cite{W}. The two models
discussed by Wang are related to known $su(4)$ and $su(1|3)$
$R$-matrices, and thus to one-dimensional chains. 
These symmetries are broken by the inclusion
of rung interactions of arbitrary strength which preserve the
integrability. 
The rung interactions appear as chemical potentials
in the equivalent one-dimensional chains, with only
a slight modification to their Bethe Ansatz solution.
On the other hand, the model introduced by Albeverio and Fei 
has rung interactions already included in   
the given interactions round an elementary face of the ladder.
The model has not been solved in general.
Those authors found the eigenvalues in the sector
of the hamiltonian in which one spin is flipped from the
ferromagnetic ground state.
Here we address how this model relates to other known models and
deduce some exact results from this equivalence.

The 2-leg ladder model is based on the symmetric $16\times 16$ 
$R$-matrix \cite{AF}
\begin{equation}
\begin{array}{l}
\check{R}(x)=\\[4mm]
\left(
\begin{array}{cccccccccccccccc}
a_1 &&&&&&&&&&&&&&&\\[2mm]
&a_3&9a_2 &&3a_2 &&&&3b_2&&&&&&&\\[2mm]
&9a_2 &a_3&&3b_2&&&&3a_2 &&&&&&&\\[2mm]
&&&a_1 &&&&&&&&&&&&\\[2mm]
&3a_2 &3b_2&&a_5&&&&a_2&&&&&&&\\[2mm]
&&&&&16x&4a_2 &&&4b_2&2a_2 &&&&&\\[2mm]
&&&&&4a_2 &a_4&&&8a_2 &4b_2&&&&&\\[2mm]
&&&&&&&a_5&&&&a_2&&3a_2&3b_2&\\[2mm]
&3b_2&3a_2 &&a_2&&&&a_5&&&&&&&\\[2mm]
&&&&&4b_2&8a_2 &&&a_4&4a_2 &&&&&\\[2mm]
&&&&&2a_2 &4b_2&&&4a_2 &16x&&&&&\\[2mm]
&&&&&&&a_2&&&&a_5&&3b_2&3a_2 &\\[2mm]
&&&&&&&&&&&&a_1 &&&\\[2mm]
&&&&&&&3a_2 &&&&3b_2&&a_3&9a_2 &\\[2mm]
&&&&&&&3b_2&&&&3a_2 &&9a_2 &a_3&\\[2mm]
&&&&&&&&&&&&&&&a_1\end{array}
\right)
\end{array}
\end{equation}
where $a_1=2( -1 + 9x )$, $a_2=-b_2=( -1 + x )$, $a_3=7 + 9x$,
$a_4=2( 3 + 5x)$, $a_5=-1 + 17x$ with $x$ arbitrary.
		      
The matrix $\check{R}(x)$ takes values in $V\otimes V$, where 
$V$ denotes a $4$-dimensional vector space.
It satisfies the Yang-Baxter equation 
\begin{equation}\label{ybe}
\check{R}_i(x)\check{R}_{i+1}(x y)\check{R}_i(y)
=\check{R}_{i+1}(y)\check{R}_i(x y)\check{R}_{i+1}(x),
\end{equation}
where $\check{R}_i$ denotes the matrix on the vector space 
$V\otimes V\otimes V$, where $\check{R}_i=\check{R} \otimes {\bf 1}$
and $\check{R}_{i+1}={\bf 1}\otimes \check{R}$ and ${\bf 1}$ is the 
identity operator on $V$. 

We see that $\check{R}$ obeys the properties
\begin{eqnarray} \check{R}(1) &=& 16\, {\bf 1} \otimes {\bf 1}, \nonumber \\
\check{R}(x) \check{R}(\ffrac{1}{x}) &=& - \ffrac{4}{x} (x-9) (9x-1)
\, {\bf 1} \otimes {\bf 1}.
\end{eqnarray}
The hamiltonian can be defined as
\begin{equation}
H_{\rm ladder} = J \sum_{i=1}^L h_i \,,
\end{equation}
where the local interactions follow from $h_i = \ffrac{1}{16} \check R'(x)$.
We find that $h$ has eigenvalues $-\ffrac{1}{8}$ (degeneracy 3) and
$\ffrac{9}{8}$ (deg. 13), in agreement with Albeverio and Fei. 
However, we disagree with the precise form of the ladder hamiltonian,
which here reads\footnote{The first two terms of the ladder
hamiltonian in Ref. \cite{AF} have co-efficient 5.}
\begin{eqnarray}\label{ladder}
H_{\rm ladder}&=& J \, \ffrac{1}{16} \displaystyle\sum_{i=1}^{L-1} [
-3\, {\bm S}_i\cdot{\bm T}_i + 13\, {\bm S}_{i+1}\cdot{\bm T}_{i+1}
+3\, ({\bm S}_i\cdot{\bm S}_{i+1}+{\bm T}_i\cdot{\bm T}_{i+1}) \nonumber\\
&&\qquad\qquad -\, 3\,
({\bm S}_i\cdot{\bm T}_{i+1} + {\bm T}_i\cdot{\bm S}_{i+1})
- 12\, ({\bm T}_i\cdot{\bm S}_{i+1})({\bm S}_i\cdot{\bm T}_{i+1}) \\
[3mm] 
&&\qquad\qquad +\, 20\, ({\bm S}_i\cdot{\bm T}_i)
({\bm S}_{i+1}\cdot{\bm T}_{i+1})
+ 12\, ({\bm S}_i\cdot{\bm S}_{i+1})({\bm T}_i\cdot{\bm T}_{i+1})]
+\ffrac{57}{4}{\bf 1}\otimes{\bf 1}]. \nonumber
\end{eqnarray}
For simplicity, we have considered open boundary conditions.
Here ${\bm S} = \ffrac{1}{2}(\sigma^x,\sigma^y,\sigma^z)$ and
${\bm T} = \ffrac{1}{2}(\sigma^x,\sigma^y,\sigma^z)$ 
are the usual spin-$\ffrac{1}{2}$ operators, with
${\bm S}_i$ and ${\bm T}_i$ spin operators on the
$i$-th rung of each leg of the ladder. 
The ladder has $L$ rungs.

\section{Hecke algebra}

Our starting point is to note that the operator $h_j$ obeys
an algebra. Specifically, if we define
\begin{equation}
U_j = \ffrac83 \left(h_j + \ffrac18 \right)  \label{scale}
\end{equation}
then $U_j$ obeys the well-known Hecke algebra, which we write here as
\begin{eqnarray}
U_j^2 &=& (q+q^{-1}) \, U_j \,,  \\
U_j U_{j+1} U_j - U_j &=& U_{j+1} U_j U_{j+1} - U_{j+1} \,, \\
\left[ U_i, U_j \right] &=& 0, \qquad \mbox{for} \quad  |i - j| > 1.
\end{eqnarray}
with $q+q^{-1} = \ffrac{10}{3}$. Thus $q=\ffrac{1}{3}$ or 3.

A number of models satisfy the Hecke algebra \cite{Chico}. 
The $4 \times 4$ representation of interest here is \cite{SZ}
\begin{equation} U_j = 
\left( \begin{array}{cccc}   q + q^{-1} &  &  &   \\
                                        & q& 1&   \\ 
                                        & 1& q^{-1}& \\   
                                        &  &  & q + q^{-1}
       \end{array}
\right) \,. \label{hecke}
\end{equation}
This is the co-product of the Casimir element belonging to the
centre of $U_q(su(2))$.
The representation (\ref{hecke}) is to be compared with the more 
well-known representation 
\begin{equation} U_j = 
\left( \begin{array}{cccc}            0 &  &  &   \\
                                        & q& 1&   \\ 
                                        & 1& q^{-1}& \\
                                        &  &  & 0
       \end{array} \label{t-l}
\right)
\end{equation}
of the Temperley-Lieb algebra.
The latter satisfies the relations (7)-(9), but with 
$U_j U_{j\pm 1} U_j - U_j =0$, and is thus a quotient of Hecke.
The representation (\ref{hecke}) may be written in terms of
spin-$\ffrac{1}{2}$ operators as
\begin{eqnarray} U_j = \ffrac12 \left( \sigma_j^x \sigma_{j+1}^x +
\sigma_j^y \sigma_{j+1}^y \right) + \ffrac14 \left( q + q^{-1} \right)
\left( \sigma_j^z \sigma_{j+1}^z - {\bf 1} \right) \nonumber\\
+ \ffrac14 \left( q^{-1} - q \right) \left(
\sigma_{j+1}^z - \sigma_j^z \right)
+ \left( q + q^{-1} \right) {\bf 1}.
\end{eqnarray}

It follows that the Hecke hamiltonian made up of spin-$\ffrac{1}{2}$ 
operators can be written
\begin{eqnarray}
H_{\rm Hecke} &=& \sum_{j=1}^{L-1} U_j  \nonumber\\
&=& \ffrac34 \left( q + q^{-1} \right) (L-1) + 
\ffrac12 \sum_{j=1}^{L-1}
\left( \sigma_j^x \sigma_{j+1}^x + \sigma_j^y \sigma_{j+1}^y
- \Delta \sigma_j^z \sigma_{j+1}^z \right) \nonumber\\ 
&& \qquad + \,\, \ffrac14 \left( q^{-1} - q \right) 
\left( \sigma_L^z - \sigma_1^z
\right) \,, \label{expr}
\end{eqnarray}
where
\begin{equation}
\Delta = - \ffrac12 \left( q + q^{-1} \right) \,.
\end{equation}
However, writing the $XXZ$ term as $H(\Delta)$, the eigenspectrum 
of (\ref{expr}) is invariant under the transformation 
$H(\Delta) = - H(-\Delta)$ and the interchange $q \leftrightarrow q^{-1}$.
This gives the eigenvalue equivalence
\begin{equation}
E_{\rm Hecke} \Leftrightarrow E_{XXZ} + 
\ffrac34 \left( q + q^{-1} \right) (L-1)\,,
\label{equiv}
\end{equation}
in which the $XXZ$ hamiltonian is defined as
\begin{equation}
H_{XXZ} = -\ffrac12 \sum_{j=1}^{L-1} 
\left( \sigma_j^x \sigma_{j+1}^x + \sigma_j^y \sigma_{j+1}^y 
+ \Delta \sigma_j^z \sigma_{j+1}^z \right) 
+ \ffrac12 \, p \left( \sigma_L^z - \sigma_1^z \right) \,, 
\label{xxz}
\end{equation}
where 
\begin{equation}
p = \ffrac12 \left( q^{-1} - q \right) \,. 
\end{equation}

This latter hamiltonian is precisely that of the open antiferromagnetic 
spin-$\ffrac{1}{2}$ $XXZ$ chain with fields
$\pm p$ at the ends of the chain. Recalling that $q=\ffrac13$,
the eigenspectrum of the open spin ladder hamiltonian (\ref{ladder}) is thus
equivalent to that of the open $XXZ$ chain with  
$\Delta = -\ffrac53$ and boundary fields $\pm \ffrac23$,
after appropriate rescaling through eq. (\ref{scale}).
In particular, for $J > 0$ the hamiltonian (\ref{ladder}) 
is equivalent to the antiferromagnetic XXZ chain, whilst
for $J < 0$ the equivalence is with the ferromagnetic XXZ chain.
For simplicity we take $J=\pm1$.

The eigenvalue equivalence (\ref{equiv}) assumes that the two
representations of the Hecke algebra are faithful, i.e. although
the representations differ in size, they share all eigenvalues in common.    
Only the multiplicity of eigenvalues differ. For given number of rungs
$L$, the ladder hamiltonian (\ref{ladder}) is of size 
$16^{L-1} \times 16^{L-1}$, whilst the equivalent $XXZ$ hamiltonian
is of size $4^{L-1} \times 4^{L-1}$. We have compared the eigenspectrum
of each hamiltonian with increasing $L$ and believe that the
representations are indeed faithful. In fact, the entire situation
is somewhat analogous to the history of the spin-1 biquadratic chain.
That model \cite{P} was mapped to the XXZ chain via the
Temperley-Lieb algebra, from which such quantities as the mass gap and
the groundstate energy etc were obtained \cite{BB}. 
These were seen to be in agreement with exact inversion relation 
calculations on the model itself \cite{K}.
The spin-1 biquadratic chain was later solved via the Bethe 
Ansatz \cite{KL}.         

From (\ref{scale}) we have 
\begin{equation}
H_{\rm ladder} = \ffrac38 H_{\rm Hecke} - \ffrac18 (L-1) \,. 
\end{equation}
On the other hand, from (\ref{equiv}) we have the eigenvalue equivalence
$
E_{\rm Hecke} \Leftrightarrow E_{XXZ} + \ffrac52 (L-1) 
$
and thus
\begin{equation}
E_{\rm ladder} \Leftrightarrow \ffrac38 E_{XXZ} + \ffrac{13}{16}(L-1)\,.
\label{rel}
\end{equation}
This is our key result.

\section{Ground state energy and mass gap}

The open $XXZ$ chain with arbitrary boundary fields has been solved
by means of the Bethe Ansatz \cite{ABBBQ}.
In particular, the solution for the case $\Delta^2 - p^2 = 1$, as
applies here, simplifies considerably.  
We shall not reproduce the equations here, but content ourselves
with recalling the relevant results. 
Consider the antiferromagnetic case first. 
In the massive region $\Delta < -1$ it is convenient to define
$q = {\rm e}^{-\theta}$.
Here the ground state energy per site, the surface free energy 
and the mass gap have all been derived \cite{BH}.\footnote{
Of course, the expressions for the ground state energy and the
mass gap are in agreement with those obtained originally \cite{OW,dG} for
periodic boundary conditions.} 
For the given XXZ normalisation, the mass gap is
\begin{eqnarray}
\Lambda_{XXZ} &=& 2 \sinh \theta \prod_{n=1}^\infty
\left( \frac{1-q^n}{1+q^n} \right)^2 \nonumber\\
&=& \ffrac{8}{3} \prod_{n=1}^\infty 
\left( \frac{3^n-1}{3^n +1} \right)^2 
= \ffrac{8}{3} \,\, 0.128108\ldots  \label{gap}
\end{eqnarray}
It thus follows from (\ref{rel}) that the ladder hamiltonian
(\ref{ladder}) has gap $\Lambda_{\rm ladder}=0.128108\ldots$ 
More generally we expect all massive excitations in the ladder
eigenspectrum to be multiples of this elementary gap.

On the other hand, the ground state energy of the open XXZ chain 
scales for large $N$ as
\begin{equation}
E_{XXZ} \sim N e_{XXZ} + f_{XXZ} \,. 
\end{equation}
The surface free energy contribution is given by
$f_{XXZ} = g - \ffrac14 \Lambda_{XXZ}$, where $g$ is a known,
though complicated, expression \cite{BH}.
The ground state energy per site is given by
\begin{equation}
e_{XXZ} = \ffrac12 \cosh \theta - \sinh \theta \left(
1 + 4 \sum_{n=1}^\infty \frac{1}{1+{\em e}^{2n\theta}} \right)\,.
\end{equation} 
It follows from (\ref{rel})  that the ground state energy per site 
of the ladder is given by
\begin{eqnarray}
e_{\rm ladder} &=& \ffrac58 - 2 \sum_{n=1}^\infty \frac{1}{1+9^n}
\nonumber\\ 
&=& 0.397527 \ldots \label{gs}
\end{eqnarray}
The surface free energy relation is
$f_{\rm ladder} = \ffrac38 f_{XXZ} - \ffrac{13}{16}$.

In the ferromagnetic regime $J < 0$ the groundstate energy of the
ladder corresponds to the trivial ferromagnetic groundstate
$\ffrac14 (q+q^{-1}) (L-1)$ of the open XXZ chain. It follows from
(\ref{rel}) that the groundstate energy of the ferromagnetic
ladder is given by $E_{\rm ladder} = - \ffrac98 (L-1)$. This value
is in agreement with the observation of Albeverio and Fei \cite{AF}.

\section{Discussion}

We have shown that a 2-leg ladder hamiltonian introduced recently by
Albeverio and Fei \cite{AF} can be made to satisfy the Hecke algebra 
for $q=\ffrac13$. As a result we have found an equivalent 
representation of the eigenspectrum in terms of the spin-$\ffrac12$
XXZ chain at $\Delta = -\ffrac53$. We considered open boundary
conditions for which the equivalent chain has surface fields 
$\pm \ffrac23$ at the ends of the chain. The values of thermodynamic
quantities such as the mass gap (\ref{gap}) and ground state energy 
per site (\ref{gs}) followed
from the known XXZ results. Periodic boundary conditions can also be
considered by imposing a twist on the periodic XXZ chain, as was done,
e.g., for the XXZ chain equivalent to the spin-1 biquadratic model \cite{BB}.

Albeverio and Fei have noted that like the well-known spin-1
Affleck-Kennedy-Lieb-Tasaki chain \cite{AKLT} the 2-leg ladder 
hamiltonian has no free parameter. However, a free parameter has
been introduced into the AKLT model via $q$-deformation \cite{BMNR,KS}.
A form of $q$-deformation should also exist for the 2-leg ladder,
corresponding to variable $q$ in the XXZ chain, again equivalent
via the Hecke algebra. However, the precise form of the 2-leg hamiltonian
would be very complicated. Nevertheless a phase transition should exist
at which the model becomes massless at the critical value $q=1$. 

Another related point is that just as Wang's 2-leg ladders can be 
generalised to $n$-leg ladders \cite{BM}, we might ask whether there
is another faithful representation of Hecke, this time of size 
$64^{L-1} \times 64^{L-1}$, corresponding to a 3-leg ladder. 
Again the hamiltonian would most likely include all possible
interactions.

Although we have seen that the 2-leg ladder hamiltonian 
provides a representation of the Hecke algebra, the
equivalence ultimately lies with the XXZ chain, and thus
with the Temperley-Lieb algebra. We thus expect that the
Hecke representation we have found here is also a quotient of Hecke. 
On another tack, considering instead the Temperley-Lieb 
representation (\ref{t-l}) with $q=\ffrac13$, we observe that as
to be expected only 
part of the eigenspectrum of the 2-leg ladder is recovered. 
Similarly if we use Saleur's Hecke representation \cite{S} 
\begin{equation} U_j =
\left( \begin{array}{cccc}   0 &  &  &   \\
                                        & q& 1&   \\
                                        & 1& q^{-1}& \\
                                        &  &  & q + q^{-1}
       \end{array}
\right) 
\end{equation}
with $q=\ffrac13$ we recover part of the ladder eigenspectrum.
The latter is known to be a quotient of Hecke \cite{ref}, but more
importantly it is free-fermionic, being equivalent to an 
XX chain \cite{S}. This gives the free-fermionic part of the
ladder eigenspectrum. As a point of further interest Saleur's 
representation can also be used to give the free-fermionic
part of the eigenspectrum of the XXZ chain.    
 
Finally we note that although the 2-leg ladder (\ref{ladder}) 
includes complicated interactions it is nevertheless a model for which 
exact results can be obtained.
It thus provides a useful testbed for numerical calculations
on more realistic models.

It is a pleasure to thank Jon Links for some helpful remarks.
This work has been supported by the Australian Research Council.




\begin{thebibliography}{99}

\def\JPA{J. Phys. A }
\def\JPC{J. Phys. C }
\def\NPB{Nucl. Phys. B }
\def\PRB{Phys. Rev. B }
\def\PRL{Phys. Rev. Lett. }
\def\RMP{Rev. Mod. Phys. }

\bibitem{KM} See, e.g., A.K. Kolezhuk and H.-J. Mikeska,
Int. J. Mod. Phys. B 12 (1998) 2325, and references therein.

\bibitem{AF} S. Albeverio and S.-M. Fei, Exactly solvable models of
generalized spin ladders, cond-mat/9807341

\bibitem{W} Y. Wang, Exact solution of a spin ladder model, cond-mat/9901168
%
\bibitem{Chico} F.C. Alcaraz, R. Koberle and A. Lima-Santos,
Int. J. Mod. Phys. A 7 (1992) 7615 

\bibitem{SZ} H. Saleur and J.-B. Zuber, in ``String Theory and
Quantum Gravity", eds M. Green et al. (World Scientific, 
Singapore, 1991) p1 

\bibitem{P} J.B. Parkinson, \JPC 21 (1988) 3793; 
J. de Physique Colloque C8 49 (1988) 1413 

\bibitem{BB} M.N. Barber and M.T. Batchelor, \PRB 40 (1989) 4621

\bibitem{K} A. Kl\"umper, Europhys. Lett. 9 (1989) 815; 
\JPA 23 (1990) 809 

\bibitem{KL} R. Koberle and A. Lima-Santos, \JPA 27 (1994) 5409 

\bibitem{ABBBQ} F.C. Alcaraz, M.N. Barber and M.T. Batchelor, 
R.J. Baxter and G.R.W. Quispel, \JPA 20 (1987) 6397

\bibitem{BH} M.T. Batchelor and C.J. Hamer, \JPA 23 (1990) 761

\bibitem{OW} R. Orbach, Phys. Rev. 112 (1958) 309;
L.R. Walker, Phys. Rev. 116 (1959) 1089

\bibitem{dG} J. des Cloizeaux and M. Gaudin, J. Math. Phys. 7 (1966) 1384

\bibitem{AKLT} I. Affleck, T. Kennedy, E.H. Lieb and H. Tasaki,
Commun. Math. Phys. 115 (1988) 477

\bibitem{BMNR} M.T. Batchelor, L. Mezincescu, R.I. Nepomechie and 
V. Rittenberg, \JPA 23 (1990) L141;
M.T. Batchelor and C.M. Yung, Int. J. Mod. Phys. B 8 (1994) 3645

\bibitem{KS} W.M. Koo and H. Saleur, Int. J. Mod. Phys. A 8 (1993) 5165

\bibitem{BM} M.T. Batchelor and M. Maslen, Exactly solvable quantum 
spin tubes and ladders, cond-mat/9907134

\bibitem{S} H. Saleur, in ``Trieste Conference on Recent Developments 
in Conformal Field Theories", eds S. Randjbar-Daemi et al. 
(World Scientific, Singapore, 1990) p160 

\bibitem{ref} P.P. Martin and V. Rittenberg, 
Int. J. Mod. Phys. A 7 (Suppl. 1B) (1992) 707

\end{thebibliography}
\end{document}